\documentclass[aip,reprint,rsi]{revtex4-1}

\usepackage{ae}

\usepackage{color}
\usepackage{bm}
\usepackage{float}
\usepackage{subfig}
\usepackage{ifthen}
\usepackage{ifpdf}
\usepackage{xspace}
\ifpdf
	\usepackage[pdftex]{graphicx}
	\usepackage[pdftex]{hyperref}
	\hypersetup{colorlinks=true,
		pdfstartview=FitV,
		linkcolor=blue,
		citecolor=blue,
		urlcolor=blue,
		pdfauthor={David Guszejnov, guszejnov@reak.bme.hu},
		pdfsubject={2012 RENATE article manuscript},
		pdftitle={2012 RENATE article manuscript}
	}
	\DeclareGraphicsExtensions{.pdf}
\else
	\usepackage[dvips]{graphicx}
	\usepackage{epsfig}
	\DeclareGraphicsExtensions{.eps}
	\usepackage{url}
\fi

 \newcommand{\renate}{{\sc renate}\xspace}

\begin{document}
 \title{Three-dimensional modeling of beam emission spectroscopy measurements in fusion plasmas} 
 \author{D.~Guszejnov}
 \email{guszejnov@reak.bme.hu}
 \affiliation{Department of Nuclear Techniques, Budapest University of Technology and Economics, Association EURATOM, H-1111 Budapest, Hungary}
 \author{G.~I.~Pokol}
 \email{pokol@reak.bme.hu}
 \affiliation{Department of Nuclear Techniques, Budapest University of Technology and Economics, Association EURATOM, H-1111 Budapest, Hungary}
 \author{I.~Pusztai}
 \affiliation{Nuclear Engineering, Applied Physics, Chalmers University of Technology, SE-412 96 G\"{o}teborg, Sweden}
 \author{D.~Refy}
 \affiliation{MTA Wigner FK RMI, Association EURATOM, Pf. 49, H-1525 Budapest, Hungary}
 \author{S.~Zoletnik}
 \affiliation{MTA Wigner FK RMI, Association EURATOM, Pf. 49, H-1525 Budapest, Hungary}
 \author{M.~Lampert}
 \affiliation{Department of Nuclear Techniques, Budapest University of Technology and Economics, Association EURATOM, H-1111 Budapest, Hungary}
 \affiliation{MTA Wigner FK RMI, Association EURATOM, Pf. 49, H-1525 Budapest, Hungary}
 \author{Y.~U.~Nam}
 \affiliation{National Fusion Research Institute, Gwahangno 113, Daejeon 305-333, Republic of Korea}

 \date{\today}
	
\begin{abstract}

One of the main diagnostic tools for measuring electron density
profiles and the characteristics of long wavelength turbulent wave structures in
fusion plasmas is Beam Emission Spectroscopy (BES). 
The increasing number of BES systems necessitated an accurate and
comprehensive simulation of BES diagnostics, which in turn motivated the development of the \renate simulation code that is the topic of this paper. \renate is a modular, fully
three-dimensional code incorporating all key features of BES systems
from the atomic physics to the observation, including an advanced
modeling of the optics. Thus \renate can be used both in the
interpretation of measured signals and the development of new BES
systems. The most important components of the code have been
successfully benchmarked against other simulation codes. The primary
results have been validated against experimental data from the KSTAR
tokamak.

\end{abstract}


\maketitle

\section{Introduction}\label{sec: intro}

Although fusion has been an intensely explored area for over 50 years,
still several key phenomena, such as the turbulent transport,
spontaneous transition to enhanced confinement regimes or the edge
localized modes, have not been adequately explained. Also a solid
predictive capability regarding to the confinement properties of
fusion plasmas is still lacking. The progress in the development and testing of theoretical models and
predictive simulation codes strongly relies on measurements from
various plasma diagnostics.  In particular there is a need for
accurate, well localized measurements of density profiles and
fluctuations, which can be accomplished by Beam Emission Spectroscopy
(BES). BES is an active plasma diagnostic which analyzes the
collisionally induced emission of either a non-intrusive Diagnostic Neutral Beam (DNB) probe or in some cases that of a Neural Beam Injection (NBI) heating beam
\cite{Thomas_BES}. The emission distribution provides information about the
distribution of density in the plasma. 

Having good spatial resolution (few cms) and high sampling
rate (few MHz) \cite{Thomas_BES} BES is ideal for measuring
density fluctuations making it the basic diagnostic tool for studying
the long wavelength fluctuations of the plasma turbulence and
different transient transport events. This resulted in a high demand
for such systems. Since determining the viability of a BES measurement configuration is not trivial, several simulation packages have been in use to aid such efforts. The emitted light intensity can be directly calculated from beam and plasma parameters if the cross sections of relevant atomic physics processes are provided. A number of codes using collisional-radiative models\cite{schweer_bes,anderson_H} have been developed which can provide the local emission intensity, such as FLYCHK \cite{FLYCHK}, NOMAD \cite{NOMAD}, SIMULA \cite{schweinzer05documentation}, NUBEAM \cite{NUBEAM}, ALCBEAM \cite{ALCBEAM} and the code developed by O. Marchuk \cite{marchuk_H}. However these are atomic physics simulation packages and accordingly the effects of the observation and beam geometry are not taken into account. From these aspects, a more detailed approach is taken in Simulation of Spectra \cite{hellermann07review}, which uses a less precise quasi-steady approximation \cite{anderson_H}, but accounts for spectral effects (e.g. motional Stark-effect) and includes basic observation and device geometry.

The development of new BES systems and the evaluation of current measurements required a detailed simulation which takes all geometrical effects, including observation, beam structure and plasma parameter distributions into account, allowing it to directly simulate measured signals. This motivated the development of the \renate
simulation code.  The aim of the current paper is to give an overview
of \renate by providing information about its capabilities and
potential applications.

The remainder of the paper is organized as follows.  In
Section~\ref{sec:renate} we describe the main structure of the code
along with the theoretical models used. Then, in
Section~\ref{sec:kstar} an example will be given for the application
of \renate along with the validation of \renate results against a
particular experimental setup on KSTAR. Finally, the results are
discussed in Sec.~\ref{sec:conclusions}.
 
\section{The \renate simulation code}\label{sec:renate}

In the present section we give an overview of the BES simulation code,
\renate (Rate Equations for Neutral Alkali-beam Technique). The code
has a modular structure and is written in the IDL language
\cite{itt10idl}. As the name suggests, the original purpose of \renate
was the modeling of alkali atomic beams (lithium and sodium)
\cite{pusztai09deconvolution}, but during the course of its development
support for the more common hydrogen isotopes was also
implemented. \renate has two different atomic physics kernels, both allowing
the modeling of the beam evolution in plasmas with mixed isotope
content and arbitrary impurity composition.

\renate is capable of calculating beam evolution by using a collisional-radiative model \cite{schweer_bes,anderson_H},
taking collisional excitation~(\textit{exc}),
de-excitation~(\textit{dexc}), ionization~(\textit{ion}), charge
exchange~(\textit{CX}) reactions and spontaneous de-excitation into
account. The recombination of the ionized beam material and the
interaction with the background electromagnetic radiation field are
neglected. This leads to the following time dependent rate equations in the frame moving with the beam atoms:
\begin{widetext}
\begin{eqnarray}
	\frac{dN_i}{dt} =  \sum_{I} n_{I}\left[ -N_{i}\left(
  \sum_{j=i+1}^{m}{R_{I}^{exc}(i \rightarrow j)}+
  \sum_{j=1}^{i-1}{R_{I}^{dexc}(i\rightarrow j)} +
  R_{I}^{ion}(i) +R_{I}^{CX}(i)\right)+
  \right. \nonumber \\
  \left. \left(\sum_{j=1}^{i-1}{N_{j}R_{I}^{exc}(j\rightarrow i)} +
  \sum_{j=i+1}^{m}{N_{j}R_{I}^{dexc}(j\rightarrow i)}\right)
  \right] - N_{i}\sum_{j=1}^{i-1}{A(i\rightarrow
  j)}+\sum_{j=i+1}^{m}{N_{j} A(j\rightarrow i)},
\label{renate_rate_equations}
\end{eqnarray}
\end{widetext}
where $N_i$ is the population of the atomic level $i$ ($i=1$ denotes the ground state and $m$ is the number of considered levels), $n_I$ is the
density of species $I$, $R$ denotes the rate coefficients and $A$
denotes the Einstein coefficients. The rate coefficients are calculated from the cross sections ($\sigma$) by assuming that the plasma species have a Maxwellian population while the beam is assumed to be monoenergetic.
\begin{equation}
\label{rate_coeff}
R=\left\langle\sigma v\right\rangle=\int f(T,\boldsymbol{v}) \left| \boldsymbol{v}-\boldsymbol{v}_b \right| \sigma\left( \left| \boldsymbol{v}-\boldsymbol{v}_b \right| \right) d\boldsymbol{v},
\end{equation}
where $\boldsymbol{v}_b$ is the beam velocity, $\boldsymbol{v}$ is the speed of the particle species interacting with the beam, while $f\left(T,\boldsymbol{v}\right)$ is their Maxwellian velocity distribution.
Equation~(\ref{renate_rate_equations}) is solved by assuming the
plasma parameters to be constant during the time needed for the beam to pass the observed region.
This allows the transformation of Eq. (\ref{renate_rate_equations}) into a space dependent system of
differential equations, which can be written in the following matrix form:
\begin{equation}
\label{renate_matrix_eq}
\frac{{d\bf{N}}}{dx}=\left[{\bf{A}}(x) n_e(x)+{\bf{B}}\right]\cdot {\bf{N}}(x)={\bf{C}}(x) \cdot {\bf{N}}(x),
\end{equation}
where the ${\bf{B}}$ matrix contains the spontaneous atomic
transitions while the ${\bf{A}}(x)$ matrix contains all collisional
processes. ${\bf{A}}$ is dependent on the spatial location $x$ through
the temperature and impurity
distribution. Due to the aforementioned transformation rule the simulation of gas jets is not feasible using Eq. (\ref{renate_matrix_eq}) as the prescribed minimum beam velocity is too high. Eq.~(\ref{renate_matrix_eq}) is solved by \renate using a
fourth order Runge-Kutta scheme.

\begin{figure*}
  \centering
  \includegraphics[width=1.0\linewidth]{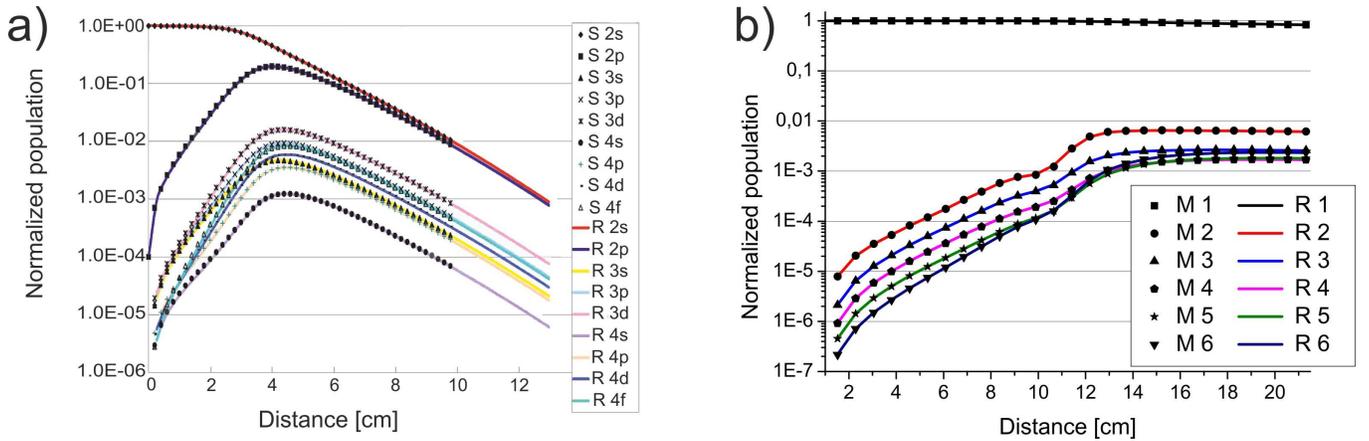}
  \caption{Evolution of the population of atomic levels along the beam normalized to
    the initial particle number. Figure (a) shows the evolution of a lithium beam calculated by \renate (R) and Simula (S). Figure (b) shows the evolution of a hydrogen beam calculated by \renate (R) and the code of O.~Marchuk (M).}
  \label{fig:bench}
\end{figure*}

In case of hydrogen isotope beams the cross sections for
$(n,l)\rightarrow (n,l\pm 1)$ transitions are so high, that a
statistical population between the sub-shells can be assumed, this is
called the bundled-n approximation \cite{bundledN} that we adopt (here $n$ and $l$ denote the principal and azimuthal quantum numbers, respectively). Due
to the fact that only a finite number of atomic levels can be taken
into account, \renate solves Eq.~(\ref{renate_rate_equations}) by
considering only the first 9 levels in case of lithium, 8 levels in
case of sodium and 6 levels (shells) in case of hydrogenic
species. This approximation is based on the fact that higher levels tend to be only lightly populated (see Fig.~\ref{fig:bench}) due to the small cross sections and high rate of spontaneous de-excitement. Numerical calculations have shown that increasing the number of considered levels above the number considered by \renate only negligibly affects the emission distribution (e.g. <1\% in case of hydrogen beams).  

The rate coefficients are calculated by the atomic physics
kernel of \renate, using atomic physics data from several sources. The
cross sections for lithium were taken from Schweinzer~et~al.~\cite{schweinzer99database}, while data for sodium were obtained
from Igenbergs~et~al.~\cite{igenbergs08database}. Cross section
data for hydrogenic species were obtained from the IAEA ALADDIN\cite{iaea10aladdin} and the Open ADAS\cite{openadas} databases with
the corrections from E.~Delabie and O.~Marchuk \cite{delabie}.

It should be noted that, unlike in the quasi-stationary model to be discussed later,
the rate coefficients for the impurity ions are scaled from the rates
of hydrogenic plasma species. Both theoretical
calculations and experiments show that the cross section of various processes
between impurities and plasma species exhibit universal behavior which is only
dependent on the charge ($q$) and energy ($E$) of the ion \cite{schweinzer99database,ryufuku80oscillatory}. The scaling
stipulates that the reduced cross section $(\sigma_{p}^{red})$ of
process $p$ at the reduced energy $(E^{red})$ is identical for all
species. The reduced cross sections and energies for excitation, charge
exchange and ionization processes are the following
\begin{eqnarray}
\sigma_{exc}^{red}=\frac{\sigma_{exc}}{q} & \qquad &E_{exc}^{red}=\frac{E_{exc}}{q}, \nonumber \\
\sigma_{CX}^{red}=\frac{\sigma_{CX}}{n^{4}_{eff}q} & \qquad  &  E_{CX}^{red}=\frac{n^{2}_{eff} E_{CX}}{q}, \nonumber \\
\sigma_{ion}^{red}=\frac{\sigma_{ion}}{n^{4}_{eff} q^{1.3}} &  \qquad & E_{ion}^{red}=\frac{n^{2}_{eff}E_{ion}}{\sqrt{q}},
\label{impurity_scaling}
\end{eqnarray}
where $n_{eff}=\sqrt{Ry/E_b}$ is the effective principal quantum number with $Ry$ as the Rydberg constant and $E_b$ as the binding energy.

The rate equation solver of \renate has been benchmarked against the
code developed by O.~Marchuk \cite{marchuk_H} for hydrogenic beams, and against the SIMULA code
\cite{schweinzer05documentation} for lithium beams
(Fig.~\ref{fig:bench}), yielding identical results within the
numerical accuracy of the calculations.

%\begin{figure*}
  %\centering
  %\subfloat[Hydrogenic benchmark]{\label{fig:H_bench}\includegraphics[width=0.5\textwidth]{bench_H.eps}}
	%\subfloat[Lithium benchmark]{\label{fig:li_bench}\includegraphics[width=0.45\textwidth]{simula.eps}}
  %\caption{Evolution of the population of atomic levels along the beam normalized to
    %the initial particle number. Figure (a) shows the evolution of a hydrogen beam calculated by \renate (R) and the code of O.~Marchuk (M). Figure (b) shows the evolution of a lithium beam calculated by \renate (R) and Simula (S)}
  %\label{fig:bench}
%\end{figure*}

Instead of solving the rate equations \renate can also utilize a simpler quasi-stationary model \cite{anderson_H},
taking atomic physics data from the Open ADAS database
\cite{openadas}. This model, that has low computational requirements,
provides a quick and approximate solution to the problem of the beam
evolution.  It assumes the life-time of excited atomic levels to be
negligible compared to the time scale of beam evolution. The beam
attenuation according to the quasi-stationary model is given by
\begin{eqnarray}
\label{beam_stopping}
N(t)=\sum_{i=1}^{\infty}{N_{i}(t)}=N_0\cdot e^{\int_{t_0}^{t}{S_{\rm eff}(\tau) d\tau}}\nonumber \\
S_{\rm eff}(\tau)=\sum_a{n_a(\tau) S_a(n_{a}(\tau),T_{a}(\tau),E)},
\end{eqnarray}
where $N(t)$ is the total number of non-ionized beam particles,
$N_{i}(t)$ is the population of the atomic level $i$, and
$S_a(n_{a},T_{a},E)$ denotes the beam stopping coefficient
corresponding to the interaction between the beam and plasma species
$a$ calculated from rate coefficients~\cite{anderson_H}
\begin{equation}
\label{stop_coeff}
S_a(n_{a},T_{a},E)=1/\left[{\bf{C}}_a^{-1}(n_{a},T_{a},E)\right ]_{1,1},
\end{equation}
using the notation of Eq.~(\ref{renate_matrix_eq}).

The emission at a given point is taken to be proportional to the
number of excited particles in the relevant atomic state and the local
density, that is
\begin{equation}
\label{qs_emission}
\lambda_{n\rightarrow n'}({\bf{r}})=\epsilon_{n\rightarrow n'}(n_{e}({\bf{r}}),T_{e}({\bf{r}}),E) \cdot n_{e}({\bf{r}})\cdot N({\bf{r}}),
\end{equation}
where $\lambda_{n\rightarrow n'}({\bf{r}})$ is the emission density
corresponding to the $n\rightarrow n'$ transition at a specific
spatial position while $\epsilon_{n\rightarrow n'}(n_{e},T_{e},E)$ is
the effective emission coefficient.
\begin{equation}
\label{emm_coeff}
\epsilon_{n\rightarrow n'}(n_{e},T_{e},E)=A_{n\rightarrow n'}\left(N_n^{(S)}/N_e N_1\right)F_n^{(1)},
\end{equation}
where $A_{n\rightarrow n'}$ is the Einstein coefficient, $N_n^{(S)}$ is the Saha-Boltzmann population for principal quantum shell $n$ and $F_n^{(1)}$ is the effective contribution of ground level excitations to the population of level $n$ \cite{anderson_H}.

Both the beam stopping and
effective emission coefficients are provided by the OpenADAS atomic
physics database \cite{openadas}. Since the quasi-stationary model
neglects the finite life-time of excited atomic levels, the results
are posteriorly smoothed by assuming exponential decay for the excited
population to account for the effect. In practice this means that the emission distribution along the beamline is convolved with an exponential decay whose characteristic length is set to be the same as the observed excited state's.

\begin{figure}
  \centering
	\includegraphics[width=1.0\linewidth]{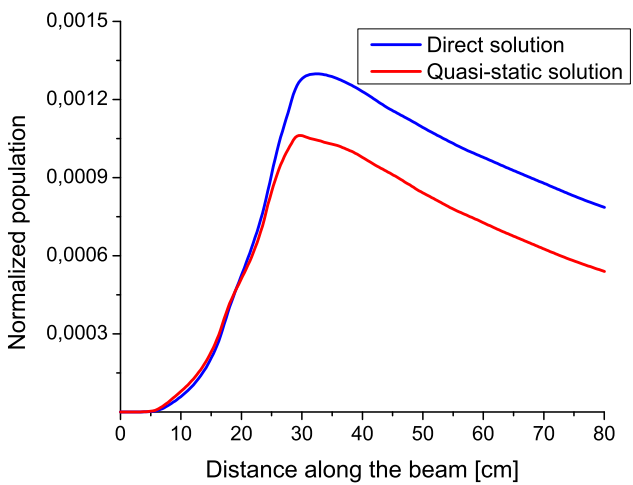}
  \caption{\label{fig:beam_evol_qs}{Evolution of the population of the observed atomic level along the beam normalized to the initial particle number for KSTAR configuration detailed in Section \ref{sec:kstar}.}}
\end{figure}

The basic assumption of the quasi-static model is that the atomic relaxation is much faster than the rate of plasma parameter change which is proportional to the beam velocity. In some cases this can be violated in the plasma pedestal leading to erroneous results (Fig. \ref{fig:beam_evol_qs}).

While there are quite a few simulation codes which can calculate beam
evolution \cite{hellermann07review, FLYCHK, NOMAD}, some even in 3D \cite{NUBEAM, ALCBEAM}, \renate is also able to take all the subtle geometrical effects of the observation into account. \renate calculates
the beam evolution along the beamline in three dimensions, while
assuming the plasma density, temperature and impurity composition to
be flux functions and distributing them according to the provided
profiles and magnetic geometry of the configuration.

\begin{figure}
  \centering
	\includegraphics[width=1.0\linewidth]{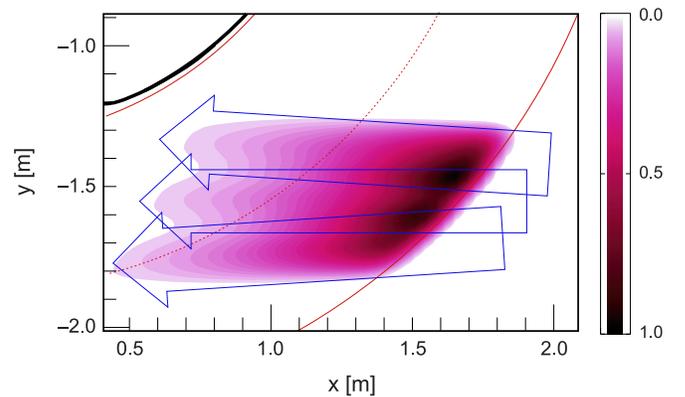}
  \caption{\label{fig:complexbeam}{Toroidal projection of the simulated relative emission intensity (purple tone color coding) of a heating beam with three ion sources (directions with blue arrows) in a hypothetical KSTAR configuration, also depicting magnetic axis (dotted red) and the edge of the plasma (red).}}
\end{figure}

As it was previously mentioned, BES diagnostic measurements are
performed not only on thin diagnostic beams but on heating beams as
well. The latter - unlike probe beams - are in general not localized to a
poloidal plane, hence the need for a full 3D simulation. Heating beams
require large and complex ion sources, thus the 3D structure of the
neutral beam is not necessarily trivial (as shown in
Fig.~\ref{fig:complexbeam}), which in turn could have a significant effect
on the measured signal. To account for this, the neutral beam
itself is modeled as a set of infinitesimally thin virtual beams, for
which the beam evolution is calculated individually in the 3D model of
the tokamak.

\begin{figure}
  \centering
	\includegraphics[width=0.5\linewidth]{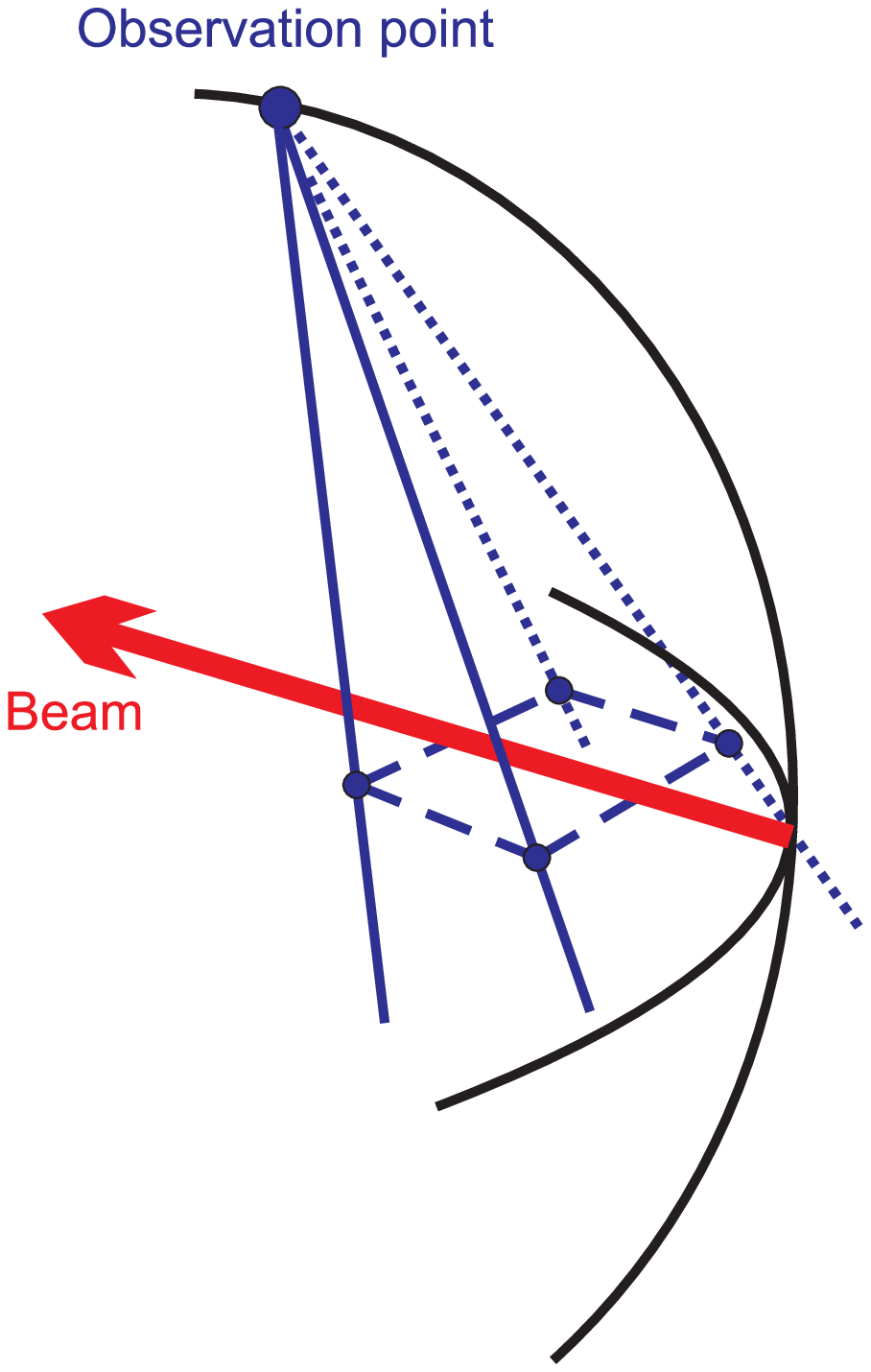}
  \caption{\label{fig:pinhole}{Schematic figure of the pinhole optical
      model of \renate showing the pyramid shaped observed area of
      space, the beamline and the observation point.}}
	
\end{figure}

Having calculated the emission distribution along each virtual beam,
contributions to the observation channels are summed up by the optical
modules. There are two possible choices for the observation modeling,
representing different levels of sophistication. The simpler module
utilizes a pinhole optics approximation which assumes that each
channel of the optical system can only detect light that was emitted within a
pyramid whose apex is the observation point, see
Fig.~\ref{fig:pinhole}. The amount of light reaching the detector
surface is assumed to be equivalent to that reaching the forward
aperture, which is calculated by integrating the emission density
weighted by the geometrical efficiency factor (coming from aperture distance, size and orientation) within the observed
volume of space.

\begin{figure}
  \centering
  \includegraphics[width=0.9\linewidth]{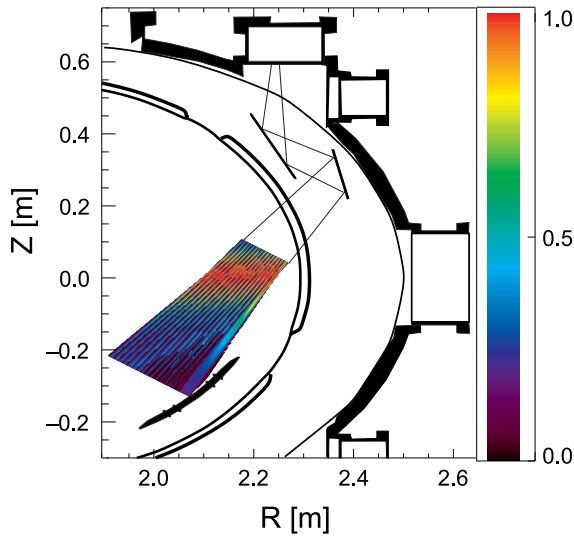}
  \caption{\label{fig:zemax} {Contour plot of the poloidal projection
      of the transfer matrices of each detector channel created by the
      advanced optical module of \renate in a hypothetical TEXTOR configuration, also denoting the structural
      elements of the device and the optical paths of the terminal
      line of sights.}}	
\end{figure}

The second, more sophisticated optical module uses ray-tracing through
the Zemax \cite{zemax} model of the optical system to calculate the observation
efficiency for every point in space, thus the transfer matrix of the
optical system (Fig.~\ref{fig:zemax}). This involves a Monte Carlo
simulation - carried out by Zemax - that creates a large number of
randomly generated rays originating from a point-like source. These
are traced through the optical system, taking the characteristics of
each optical element into account. Then the observation efficiency is
simply the light intensity reaching the individual detector surface
divided by the emitted intensity.

This method provides more accurate results, as it takes all the optical
elements of the system into account, however it can only be used in
the final design phase of BES systems when the detailed optical plan is
available. Meanwhile the pinhole optics module is much more flexible
allowing the tryout of a multitude of different configurations without
the computationally demanding calculation of transfer matrices. It
should be noted that while \renate can calculate the Doppler-shifted
spectrum and thus take filtering into account, it is currently unable
to take other wavelength shifting effects (e.g. motional Stark-effect)
into account, which means that filtering efficiency for each channel has to be provided
externally (e.g. by another simulation code, like Simulation of Spectra \cite{hellermann07review}).

\begin{figure}
\begin{center}
\includegraphics[width=0.95 \linewidth]{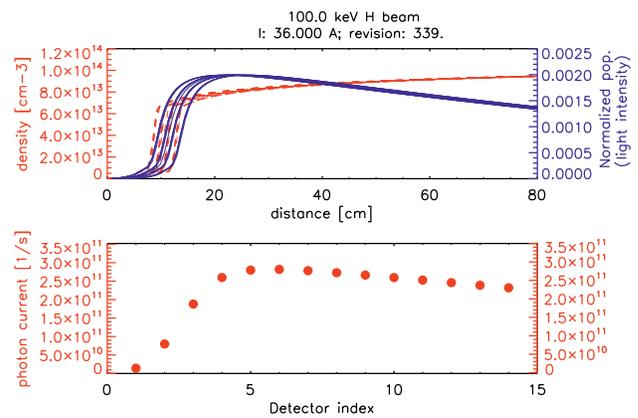}
\end{center}
\caption{\label{fig:ext_iter}{Graphical output  of \renate showing the
    density along the beamlets (top-red), light profiles (top-blue)
    and the photon currents reaching the detector surfaces (bottom)}}
\end{figure}

The final result of the calculation is the expected absolute photon
currents ($1/\rm{s}$) reaching the surface of the individual detector
segments (Fig.~\ref{fig:ext_iter}).

To give a complete account of the capabilities of \renate, we note that
it can compute the density perturbation response matrices, which in
turn could be used to determine the response in detected signal of the
investigated BES system to arbitrary density perturbations. Using this
matrix, the plasma density fluctuations can be converted into signal fluctuations, however this is only possible if
the response is proportional to the strength of the
perturbation. It is also possible to try and calculate the density fluctuations for a given BES system from the experimental signal data. This
aspect of the \renate simulation code will be thoroughly explored in a
follow up article.

\section{Application to KSTAR trial measurement}\label{sec:kstar}

\begin{figure*}
\begin{center}
\includegraphics[width=0.75 \linewidth]{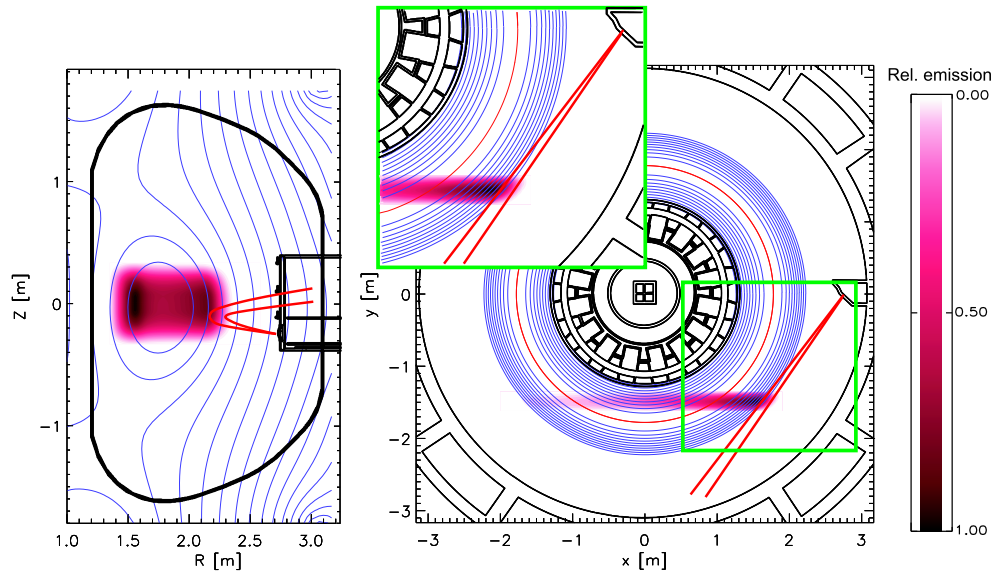}
\end{center}
\caption{\label{fig:kstar}{Toroidal and poloidal projections of the
    simulated configuration for KSTAR BES, showing some structural
    elements (black), the flux surfaces (blue), the magnetic axis (red circle), the relative emission
    density (purple tone color coding) and the edge lines of sights (red)
    from the observation point}}
\end{figure*}

The first demonstration of the capabilities of \renate includes
applications to the KSTAR BES system \cite{guszejnov11kstar}.  With
the help of \renate the photon currents reaching the detector surfaces
could be calculated, which - combined with the given detector type's
characteristics - enabled the calculation of the signal-to-noise ratio
for each detector segment. \renate was also capable of providing
estimates of the 3D spatial and 2D poloidal resolution of the
fluctuation BES system, by calculating the the response matrix and
integrating it along magnetic field lines. The poloidal and toroidal
projections of the simulated configuration are depicted on
Fig.~\ref{fig:kstar}. 

\begin{figure}
\begin{center}
\includegraphics[width=1.0 \linewidth]{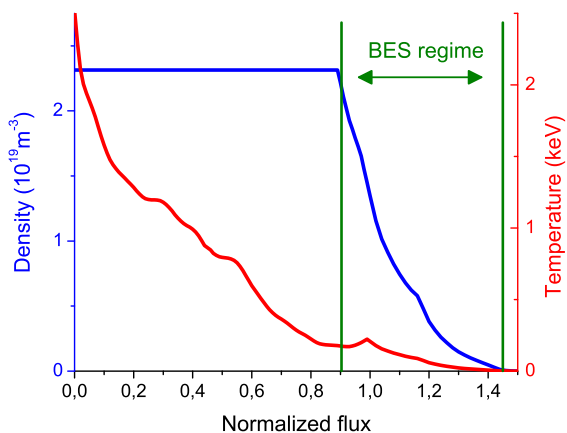}
\end{center}
\caption{\label{fig:profiles}{Plasma parameter profiles used in the BES modeling of the KSTAR BES trial measurement. Density data (blue) were fitted to interferometry results while temperature data (red) were obtained from ECE measurements.}}
\end{figure}

Using measured data the simulations have been rerun for the parameters of KSTAR discharge \#6123 at 1744 ms, and its results have been compared against experimental
results from the KSTAR BES trial measurement \cite{yongunnamKSTAR}. The magnetic geometry was reconstructed using EFIT \cite{EFIT} while the temperature profile was obtained from Electron Cyclotron Emission (ECE) measurements, thus assuming $T_i=T_e$ (Fig. \ref{fig:profiles}). Since at the time of the measurements KSTAR had only a single interferometry channel for plasma density measurement, it was not possible to get localized density data from the experiment. Instead a profile of specific shape was fitted to the line integrated data. This density profile is flat in the core plasma, while follows the shape of the relative BES emission in the edge as shown in Fig. \ref{fig:profiles}. Impurity data were also unavailable thus a homogeneous 1\% carbon impurity was assumed, corresponding to $Z_{eff}=1.35$.

\begin{figure}
\begin{center}
\includegraphics[width=1.0 \linewidth]{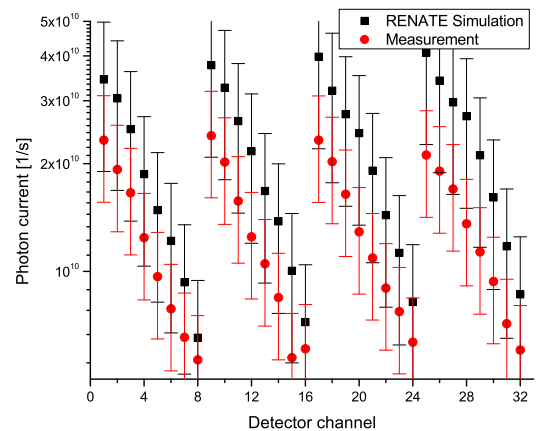}
\end{center}
\caption{\label{fig:absolute_photon}{Measured and simulated photon currents for individual detector channels for KSTAR \#6123 at 1744 ms. The detectors are sorted into a 4 x 8 matrix, indexing start from the detector of the first row observing the innermost part of the plasma. The error estimates are high due to inaccurate input data and experimental uncertainties.}}
\end{figure}

During the experiment only the central beam source was active, resulting in a single deuterium beam ($13\,\rm{A}$, $90\,\rm{keV}$) with known current distribution. The measurement was carried out using a 4 x 8 matrix of detectors with spatial resolution of $1\,\rm{cm}$. Subtracting the background light from the measured data and taking into account the transmittance of the optical system along with the effect of the filtering and detector amplification yielded the absolute photon current reaching the aperture for each individual channels. This method however, carries a significant relative error mainly because only a relative calibration of the detector amplification was possible from gas shots. By estimating this error term to be roughly 30\% and assuming the other terms to be independent, the measured data have a total 33\% relative error. These data can be compared against \renate results as shown in Fig. \ref{fig:absolute_photon}. The density profile of this simulation was highly arbitrary and a basic pinhole optics module was used to model the observation, which represent error terms of 40\% and 20\% respectively. This means an estimated relative error of 44\% for the the simulation results. 

\begin{figure}
\begin{center}
\includegraphics[width=1.0 \linewidth]{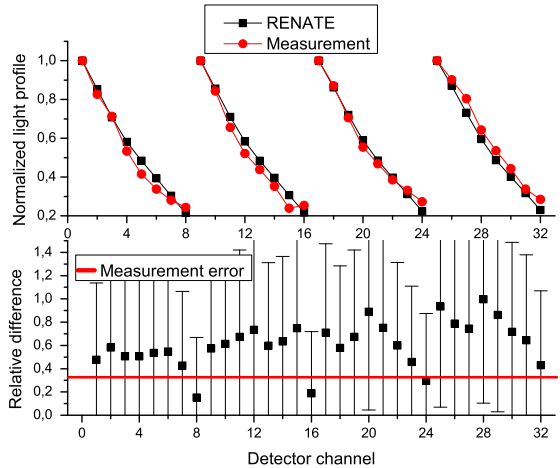}
\end{center}
\caption{\label{fig:trans_profiles}{Normalized light profiles of the individual detector rows (upper figure) and the relative error of the simulation compared to the measurement (lower figure).}}
\end{figure}

As shown in Fig. \ref{fig:absolute_photon} the simulated signal is close to the measurement. The upper part of Fig. \ref{fig:trans_profiles} shows the normalized light profiles of the individual detector rows, where it is evident that the shape of the simulated signal matches that of the measurement. This is no surprise as the density profile was fitted to the relative BES emission which is roughly proportional to the local density as shown by Equation (\ref{renate_matrix_eq}). The real test of the code is the lower part of Fig. \ref{fig:trans_profiles}, which shows the relative error of the simulation compared to the measurement. 

On average the simulated results are 60\% higher than the measurement. Considering that the standard deviation of the error is rather small (20\%) the dominant term is caused by a systematic error which can be caused by a number of effects. 
\begin{itemize}
	\item The density profile is the main source of error as it is the primary input of \renate simulations. Since only line integrated density was available, the shape of the density profile in the non-BES region of the plasma could cause a systematic error. Using a parabolic profile for the plasma center instead of the flat one shown in Fig. \ref{fig:profiles} could mitigate this effect. It was found that a 40\% reduction of the BES region density eliminates this term leaving only an error of only 20\%, which could easily be attributed to other uncertainties (e.g. pinhole optics model).
	\item The \renate simulation utilized a simple pinhole optical model which could have easily neglected a portion of the real observed area.
	\item The atomic physics data of most hydrogenic transitions have a relative error of 10-20\% \cite{iaea10aladdin, openadas}.
\end{itemize}

\section{Conclusions}\label{sec:conclusions}
This paper presents an overview of the \renate code, which is designed
to help the development of new BES systems and the evaluation of
experimental data. It is a modular BES simulation capable of simulating most key levels of a BES system. It
possesses two different types of atomic physics kernel, allowing it to
calculate beam evolution either by solving the rate equations or by
employing a quasi-steady approximation. Both of these methods utilize the latest
atomic physics data. \renate is able to model the three-dimensional geometry of the simulated system in detail,
including the inner structure of the beam and the optical setup,
allowing it to model complete BES systems, in particular heating beam measurements.

The rate equation solver module of \renate has been benchmarked for
hydrogenic species and lithium. Meanwhile the experimental results for the
KSTAR BES trial measurement agree with the simulated results well within the estimated error. In the future \renate is to be compared against
other experimental results and is to be upgraded to be able to take
more wavelength shifting effects (e.g. motional Stark-effect) into
account.

\begin{acknowledgments}
This work was partly funded by the European Communities under
  Association Contract between EURATOM and HAS. The views and opinions
  expressed herein do not necessarily reflect those of the European
  Commission. Authors of this paper also acknowledge the support of the Hungarian National Development Agency under Contract No. NAP-1-2005-0013.
	We are grateful for the support of E. Delabie, O. Marchuk and J. Schweinzer.
\end{acknowledgments}

%\bibliography{hivatkozas}

\begin{thebibliography}{10}

\bibitem{Thomas_BES}
D.~M. Thomas, G.~R. McKee, K.~H. Burrell, F.~Levinton, E.~L. Foley, and R.~K.
  Fisher.
%\newblock Active spectroscopy.
\newblock {\em Fusion Science and Technology}, 53(2):488 -- 527, 2008.

\bibitem{schweer_bes}
B.~Scwheer.
%\newblock {Application of Atomic Beams in Combination with Spectroscopic
%  Observation for Plasma Diagnostic}.
\newblock {\em Fusion Science and Technology}, 53:425--432, 2008.

\bibitem{anderson_H}
H.~Anderson, M.~G. von Hellermann, R.~Hoekstra, L.~D. Horton, A.~C. Howman,
  R.~W.~T. Konig, R.~Martin, R.~E. Olson, and H.~P. Summers.
%\newblock Neutral beam stopping and emission in fusion plasmas i: deuterium
%  beams.
\newblock {\em Plasma Physics and Controlled Fusion}, 42(7):781, 2009.

\bibitem{FLYCHK}
H.-K. Chung, M.H. Chen, W.L. Morgan, Y.~Ralchenko, and R.W. Lee.
%\newblock {FLYCHK}: Generalized population kinetics and spectral model for
%  rapid spectroscopic analysis for all elements.
\newblock {\em High Energy Density Physics}, 1(1):3 -- 12, 2005.

\bibitem{NOMAD}
Yuri~V. Ralchenko and Yitzhak Maron.
%\newblock Accelerated recombination due to resonant deexcitation of metastable
%  states.
\newblock {\em Journal of Quantitative Spectroscopy and Radiative Transfer},
  71:609 -- 621, 2001.

\bibitem{schweinzer05documentation}
J.~Schweinzer.
\newblock {\em {Documentation on the Installation of a Code Package at {NIFS}
  for Reconstructing Density Profiles from Lithium Beam Emission Data}}.
\newblock MPI f\"{u}r Plasmaphysik, Garching, 2005.

\bibitem{NUBEAM}
Alexei Pankin, Douglas McCune, Robert Andre, Glenn Bateman, and Arnold Kritz.
%\newblock The tokamak monte carlo fast ion module nubeam in the national
%  transport code collaboration library.
\newblock {\em Computer Physics Communications}, 159(3):157 -- 184, 2004.

\bibitem{ALCBEAM}
I.O. Bespamyatnov, W.L. Rowan, and K.T. Liao.
%\newblock Alcbeam – neutral beam formation and propagation code for
  %beam-based plasma diagnostics.
\newblock {\em Computer Physics Communications}, 183(3):669 -- 676, 2012.

\bibitem{marchuk_H}
O.~Marchuk, G.~Bertschinger, W.~Biel, E.~Delabie, M.~G. von Hellermann,
  R.~Jaspers, and D.~Reiter.
%\newblock Review of atomic data needs for active charge-exchange spectroscopy
  %on iter.
\newblock {\em Review of Scientific Instruments}, 79:10F532, 2008.

\bibitem{hellermann07review}
M.G. von Hellermann, R.~Jaspers, W.~Biel, O.~Neubauer, N.~Hawkes, Y.~Kaschuck,
  V.~Serov, S.~Tugarinov, D.~Thomas, and W.~Vliegenthart.
%\newblock {Review of Beam Aided Diagnostics for ITER}.
\newblock In {\em {Proceedings of the 21st IAEA Conference, Chengdu, 16-21
  October 2006}}, volume IAEA-CN-149 of {\em IAEA Conference and Symposium
  Papers}, pages {IT/P1--26}. IAEA, Vienna, 2007.

\bibitem{itt10idl}
{ITT} Data~Visualization Solutions.
\newblock {\em {{IDL} official webpage}}.
\newblock http://www.ittvis.com/ProductServices/IDL.aspx, May 2010.

\bibitem{pusztai09deconvolution}
I.~Pusztai, G.~Pokol, D.~Dunai, D.~R\'{e}fy, G.~P\'{o}r, G.~Anda, S.~Zoletnik,
  and J.~Schweinzer.
%\newblock Deconvolution-based correction of alkali beam emission spectroscopy
  %density profile measurements.
\newblock {\em Review of Scientific Instruments}, 80(8):083502, 2009.

\bibitem{bundledN}
A.~Burgess and H.~P. Summers.
%\newblock {The recombination and level populations of ions. I - Hydrogen and
  %hydrogenic ions}.
\newblock {\em Monthly Notices of the Royal Astronomical Society}, 174:345 --
  391, 1976.

\bibitem{schweinzer99database}
J.~Schweinzer, R.~Brandenburg, I.~Bray, R.~Hoekstra, F.~Aumayr, R.K. Janev, and
  HP. Winter.
%\newblock {Database for inelastic collisions of lithium atoms with electrons,
  %protons, and multiply charged ions}.
\newblock {\em Atomic Data and Nuclear Data Tables}, 72(2):239 -- 273, 1999.

\bibitem{igenbergs08database}
K.~Igenbergs, J.~Schweinzer, I.~Bray, D.~Bridi, and F.~Aumayr.
%\newblock {Database for inelastic collisions of sodium atoms with electrons,
  %protons, and multiply charged ions}.
\newblock {\em Atomic Data and Nuclear Data Tables}, 94(6):981 -- 1014, 2008.

\bibitem{iaea10aladdin}
IAEA.
\newblock {\em {{ALADDIN}}}.
\newblock http://www-amdis.iaea.org/ALADDIN/, May 2010.

\bibitem{openadas}
{ADAS Project}.
\newblock {\em {{Open ADAS}}}.
\newblock http://open.adas.ac.uk, January 2011.

\bibitem{delabie}
E~Delabie, M~Brix, C~Giroud, R~J~E Jaspers, O~Marchuk, M~G O'Mullane,
  Yu~Ralchenko, E~Surrey, M~G von Hellermann, K~D Zastrow, and JET-EFDA
  Contributors.
%\newblock Consistency of atomic data for the interpretation of beam emission
  %spectra.
\newblock {\em Plasma Physics and Controlled Fusion}, 52(12):125008, 2010.

\bibitem{ryufuku80oscillatory}
H.~Ryufuku, K.~Sasaki, and T.~Watanabe.
%\newblock Oscillatory behavior of charge transfer cross sections as a function
  %of the charge of projectiles in low-energy collisions.
\newblock {\em Phys. Rev. A}, 21(3):745--750, Mar 1980.

\bibitem{zemax}
Warren~J. Smith.
\newblock {\em {Modern Optical Engineering}}.
\newblock McGraw-Hill., 4 edition, 2007.

\bibitem{guszejnov11kstar}
D.~Guszejnov and G.~I. Pokol.
\newblock {First Results of RENATE Simulation for the KSTAR Tokamak}.
\newblock In {\em Joint Workshop for KSTAR BES Diagnostics (Budapest)}, 2011.

\bibitem{yongunnamKSTAR}
Y.~U. Nam, S.~Zoletnik, M.~Lampert, and \'{A}. Kov\'{a}csik.
%\newblock Analysis of edge density fluctuation measured by trial kstar beam
  %emission spectroscopy system.
\newblock {\em Review of Scientific Instruments}, 83(10):10D531, 2012.

\bibitem{EFIT}
L.L. Lao, J.R. Ferron, R.J. Groebner, W.~Howl, H.~St. John, E.J. Strait, and
  T.S. Taylor.
%\newblock Equilibrium analysis of current profiles in tokamaks.
\newblock {\em Nuclear Fusion}, 30(6):1035, 1990.

\end{thebibliography}
%\bibliographystyle{unsrt}

\end{document}